\documentclass{ws-ijmpa}
\usepackage[super,compress]{cite}
\usepackage{graphicx}

\begin{document}
\markboth{S.~Cebri\'an}{Cosmogenic activation of materials}

%
\catchline{}{}{}{}{}
%

\title{Cosmogenic activation of materials}

\author{Susana Cebri\'an}

\address{Grupo de F\'isica Nuclear y Astropart\'iculas, \\Universidad de Zaragoza, Calle Pedro Cerbuna 12, 50009 Zaragoza, Spain\\
Laboratorio Subterr\'aneo de Canfranc, Paseo de los Ayerbe s/n \\
22880 Canfranc Estaci\'on, Huesca, Spain \\
scebrian@unizar.es}

\maketitle

\begin{history}
\received{Day Month Year}
\revised{Day Month Year}
\end{history}

\begin{abstract}
Experiments looking for rare events like the direct detection of
dark matter particles, neutrino interactions or the nuclear double
beta decay are operated deep underground to suppress the effect of
cosmic rays. But the production of radioactive isotopes in materials
due to previous exposure to cosmic rays is an hazard when ultra-low
background conditions are required. In this context, the generation
of long-lived products by cosmic nucleons has been studied for many
detector media and for other materials commonly used. Here, the main
results obtained on the quantification of activation yields on the
Earth's surface will be summarized, considering both measurements
and calculations following different approaches. The isotope
production cross sections and the cosmic ray spectrum are the two
main ingredients when calculating this cosmogenic activation; the
different alternatives for implementing them will be discussed.
Activation that can take place deep underground mainly due to cosmic
muons will be briefly commented too. Presently, the experimental
results for the cosmogenic production of radioisotopes are scarce
and discrepancies between different calculations are important in
many cases, but the increasing interest on this background source
which is becoming more and more relevant can help to change this
situation.

\end{abstract}

\ccode{PACS numbers: 13.85.Tp; 23.40.-s; 25.40.-h; 25.30.Mr;
95.35.+d}

\section{Introduction}

Experiments searching for rare phenomena like the interaction of
Weakly Interacting Massive Particles (WIMPs) which could be filling
the galactic dark matter halo, the detection of elusive neutrinos or
the nuclear double beta decay of some nuclei require detectors
working in ultra-low background conditions and taking data for very
long periods of time at the scale of a few years due to the
extremely low counting rates expected. Operating in deep underground
locations, using active and passive shields and selecting carefully
radiopure materials reduce very efficiently the background for this
kind of experiments \cite{heusser,formaggio}. In this context,
long-lived radioactive impurities in the materials of the set-up
induced by the exposure to cosmic rays at sea level (during
fabrication, transport and storage) may be even more important than
residual contamination from primordial radionuclides and become very
problematic. For instance, the poor knowledge of cosmic ray
activation in detector materials is highlighted in
Ref.~\citen{gondolo} as one of the three main uncertain nuclear
physics aspects of relevance in the direct detection approach
pursued to solve the dark matter problem. Production of radioactive
isotopes by exposure to cosmic rays is a well known issue
\cite{lal,beer} and several cosmogenic isotopes are important in
extraterrestrial and terrestrial studies in many different fields,
dealing with astrophysics, geophysics, paleontology and archaeology.

One of the most relevant processes in the cosmogenic production of
isotopes is the spallation of nuclei by high energy nucleons, but
other reactions like fragmentation, induced fission or capture can
be very important for some nuclei. Since protons are absorbed by the
atmosphere, nuclide production is mainly dominated by neutrons at
the Earth's surface, but if materials are flown at high altitudes,
in addition to the fact that the cosmic flux is much greater,
protons will produce significant activation as well.

In principle, cosmogenic activation can be kept under control by
minimizing exposure at surface and storing materials underground,
avoiding flights and even using shields against the hadronic
component of cosmic rays during surface detector building or
operation. In addition, purification techniques could help reducing
the produced background isotopes. But since these requirements
usually complicate the preparation of experiments (for example,
while crystal growth and detector mounting steps) it would be
desirable to have reliable tools to quantify the real danger of
exposing the different materials to cosmic rays. Production rates of
cosmogenic isotopes in all the materials present in the experimental
set-up, as well as the corresponding cosmic rays exposure history,
must be both well known in order to assess the relevance of this
effect in the achievable sensitivity of an experiment. Direct
measurements, by screening of exposed materials in very low
background conditions as those achieved in underground laboratories,
and calculations of production rates and yields, following different
approaches, have been made for several materials in the context of
dark matter, double beta decay and neutrino experiments. Many
different studies are available for germanium and interesting
results have been derived in the last years also for other detector
media like tellurium and tellurium oxide, sodium iodide, xenon,
argon or neodymium as well as for materials commonly used in the
set-ups like copper, lead, stainless steel or titanium (see for
instance the summaries in
Refs.~\citen{cebrianlrt2013,kudrylrt2017}). However, systematic
studies for other targets are missed.

The relevant long-lived radioactive isotopes cosmogenically induced
are in general different for each target material, although there
are some common dangerous products like tritium, because being a spallation product, it can be generated in any material. Tritium is
specially relevant for dark matter experiments for its decay
properties (it is a pure beta emitter with transition energy of
18.591~keV and a long half-life of 12.312~y \cite{DDEPH3}) when
induced in the detector medium. Quantification of tritium cosmogenic
production is not easy, neither experimentally since its beta
emissions are hard to disentangle from other background
contributions, nor by calculations, as tritium can be produced by
different reaction channels.

The aim of this work is to summarize and compare the main results
obtained for estimating the activation yields of relevant long-lived
radioactive isotopes due to cosmic nucleons on the Earth's surface
in the materials of interest for rare event experiments, taken into
consideration both measurements and calculations. The structure of
the paper is as follows. First, section~\ref{cal} describes the
relevant calculation tools available. Activation results derived for
different detector media are summarized in the following sections,
including semiconductor materials like germanium
(section~\ref{secge}) and silicon (section~\ref{secsi}), tellurium
used in different detectors (section~\ref{secte}), scintillators
like sodium iodide (section~\ref{secnai}) and noble liquid-gas
detectors as xenon (section~\ref{secxe}) and argon
(section~\ref{secar}). The production of cosmogenic isotopes in
other materials typically used in experiments is also considered, in
particular, for copper (section~\ref{seccu}), lead
(section~\ref{secpb}), stainless steel (section~\ref{secss}) and
titanium (section~\ref{secti}). Results for other targets are
collected in section~\ref{other}. For each material, the available
estimates of production rates of the relevant radionuclides will be
presented, emphasizing the comparison with experimental results
whenever possible. Some results concerning the activation of
materials underground are mentioned too in section~\ref{under}
before the final summary.

\section{Calculation Tools}
\label{cal}

The activity $A$ of a particular isotope with decay constant
$\lambda$ induced in a material must be evaluated taking into
account the time of exposure to cosmic rays  ($t_{exp}$), the cooling time corresponding to the time spent underground and sheltered from cosmic rays ($t_{cool}$) and the production rate $R$:

\begin{equation}
A = R [1-\exp(-\lambda t_{exp})] \exp(-\lambda t_{cool}).
\end{equation}

\noindent As it will be shown in the next sections, there are some
direct measurements of productions rates in some materials as the
saturation activity, but unfortunately they are not very common.
Consequently, in many cases production rates have to be estimated
from two basic energy-dependent ingredients, the flux $\phi$ of
cosmic rays and the cross-section $\sigma$ of isotope production:

\begin{equation}
R=N\int\sigma(E)\phi(E)dE \label{eqrate}.
\end{equation}

\noindent $N$ being the number of target nuclei and $E$ the particle
energy.

The excitation function for the production of a certain isotope by
nucleons in a target over a wide range of energies (typically from
some MeV up to several GeV) can be hardly obtained experimentally,
since the measurements of production cross-sections with beams are
long, expensive and there are not many available facilities to carry
them out. The use of computational codes to complete information on
the excitation functions is therefore mandatory. In addition,
measurements are usually performed on targets with the natural
composition of isotopes for a given element, often determining only
cumulative yields of residual nuclei. Reliable calculations are
required to provide independent yields for isotopically separated
targets if necessary. But in any case, experimental data are
essential to check the reliability of calculations. The suitability
of a code to a particular activation problem depends on energy,
targets and projectiles to be considered. Some systematic and
extensive comparisons of calculations and available measurements
have been made based on analysis of deviation factors $F$, defined
as:
\begin{equation}
F=10^{\sqrt{d}},\hspace{0.5cm}
d=\frac{1}{n}\sum_{i}(\log\sigma_{exp,i}-\log\sigma_{cal,i})^{2}
\label{eqF}
\end{equation}
\noindent $n$ being the number of pairs of experimental and
calculated cross sections $\sigma_{exp,i}$ and $\sigma_{cal,i}$ at
the same energies.

There are several possibilities to obtain values of {\bf production
cross sections}:

\begin{itemlist}
\item Experimental results can be found at EXFOR database (CSISRS in
USA) \cite{exfor}, an extensive compilation of nuclear reaction data
from thousands of experiments. Available data for a particular
target, projectile, energy or reaction can be easily searched for by
means of a public web form. NUCLEX \cite{nuclex} is also a
compilation of experimental data.

\item Semiempirical formulae were deduced for nucleon-nucleus cross sections for different reactions
(break-up, spallation, fission,\dots) exploiting systematic
regularities and tuning parameters to best fit available
experimental results. The famous Silberberg\&Tsao equations
\cite{tsao1,tsao2,tsao2i,tsao2ii,tsao3} can be applied for light and
heavy targets (A$\geq$3), for a wide range of product radionuclides
(A$\geq$6) and at energies above 100~MeV. These equations have been
implemented in different codes, which offer very fast calculations
in contrast to Monte Carlo simulations:
\begin{itemlist}
\item COSMO \cite{cosmo} is a FORTRAN program with three modes of calculation:
excitation curve of a nuclide for a specified target, mass yield
curve for given target and energy and final activities produced for
a target exposed to cosmic rays. It allows a complete treatment for
targets with A$<$210 and Z$<$83.
\item YIELDX \cite{tsao3} is a FORTRAN routine to calculate the production cross-section of a
nuclide in a particular target at a certain energy. It includes the
latest updates of the Silberberg \& Tsao equations.
\item ACTIVIA \cite{activia} is C++ computer package to calculate target-product cross
sections as well as production and decay yields from cosmic ray
activation. It uses semiempirical formulae but also experimental
data tables if available.
\end{itemlist}
The main limitation of the formulae is that they are based only on
proton-induced reactions; therefore, the fact that cross sections
are equal for neutrons and protons has to be assumed.

\item Monte Carlo simulation of the hadronic interaction between nucleons and
nuclei can output also production cross sections. Modeling isotope
production includes a wide range of reactions: the formation and
decay of a long-lived compound nucleus at low energies, while in the
GeV range the intranuclear cascade (INC) of nucleon-nucleon
interactions is followed by different deexcitation processes
(spallation, fragmentation, fission,\dots). Many different models
and codes implementing them have been developed in very different
contexts (studies of comic rays and astrophysics, transmutation of
nuclear waste or production of medical radioisotopes for instance):
BERTINI, ISABEL, LAHET, GEM, TALYS, HMS-ALICE, INUCL, LAQGSM, CEM,
ABLA, CASCADE, MARS, SHIELD,\dots are the names of just a few of
them. Some of these models have been integrated in general-purpose
codes like GEANT4 \cite{geant4}, FLUKA \cite{fluka} and MCNP
\cite{mcnp}. In Ref.~\citen{aguayo}, the capabilities of GEANT4 and
MCNPX to simulate neutron spallation were studied; in particular,
the neutron multiplicity predicted was compared with measurements
finding some discrepancies for low density materials. In
Ref.~\citen{head2009} preferred models for each target mass range
are selected, for neutron and proton-induced reactions while in
Ref.~\citen{titarenko} it is concluded that versions of CEM03 and
INCL+ABLA codes can be considered as the most accurate. A relevant
advantage of Monte Carlo codes is that they are applicable not only
to proton-induced but also to neutron-induced nuclear reactions.

\item Several libraries of production cross sections have been prepared combining calculations and experimental data, with different coverage of energies, projectiles and
reactions:
\begin{itemlist}
\item RNAL (Reference Neutron Activation Library) \cite{rnal} is restricted to 255
reactions.
\item LA150 \cite{la150} contains results up to 150~MeV, independently for neutrons and
protons as projectiles, using calculations from HMS-ALICE.
\item MENDL-2 and MENDL-2P (Medium Energy Nuclear Data Library) \cite{mendl} are based
on calculations using codes of the ALICE family, containing
excitation functions which cover a very wide range of target and
product nuclides, either for neutrons or protons, for energies up to
100~MeV for neutrons (MENDL-2, using ALICE92) and 200~MeV for
protons (MENDL-2P, using ALICEIPPE).
\item  TENDL (TALYS-based Evaluated Nuclear Data Library)
\cite{tendl} offers cross sections obtained with the TALYS nuclear
model code system for neutrons and protons up to 200~MeV for all the
targets.
\item An evaluated library for neutrons and protons to 1.7~GeV was
presented in Ref.~\citen{study}. It gives excitation functions
including available experimental data and calculated results using
CEM95, LAHET, and HMS-ALICE codes for selected targets and products.
\item HEAD-2009 (High Energy Activation Data) \cite{head2009} is a complete
compilation of data for neutrons and protons from 150~MeV up to
1~GeV. The choice of models was dictated by an extensive comparison
with EXFOR data. In particular, in HEPAD-2008 (High-Energy Proton
Activation Data) sub-library cross sections are obtained using a
selection of models and codes like CEM 03.01, CASCADE/INPE and
MCNPX~2.6.0. Only targets with Z$\geq$12 are considered.

The detailed studies of excitation functions performed in different
areas, like Positron Emission Tomography (PET) nuclear medicine
imaging process, for different targets and products and based on
irradiation experiments and several types of calculations (see for
example recent works from Refs.~\citen{ef1,ef2,ef3,ef4,ef5,ef6}),
can serve as a reference to select the most adequate libraries or
packages.
\end{itemlist}
\end{itemlist}

Relevant activation processes on the Earth's surface are induced
mainly by nucleons at MeV-GeV energy range.  At sea level, the flux
of neutrons and protons is virtually the same at energies of a few
GeV; however, at lower energies the proton to neutron ratio
decreases significantly because of the absorption of charged
particles in the atmosphere. For example, at 100~MeV this ratio is
about 3\%\cite{lal}. Consequently, neutrons dominate the
contribution to the {\bf cosmic ray spectrum} to be considered at
sea level. But it must be kept in mind that proton activation, being
much smaller, is not completely negligible. The typical contribution
from protons to isotope production is quoted as $\sim$10\% of the
total in Ref.~\citen{lal}, which agrees for instance with the
results in Refs.~\citen{barabanov,wei} where proton activation in
germanium was specifically evaluated for some isotopes. Concerning
activation by other cosmic particles like muons, it is even smaller,
as confirmed by calculations in Refs.~\citen{mei2016,wei}.

\begin{figure}
 \includegraphics[height=.5\textheight]{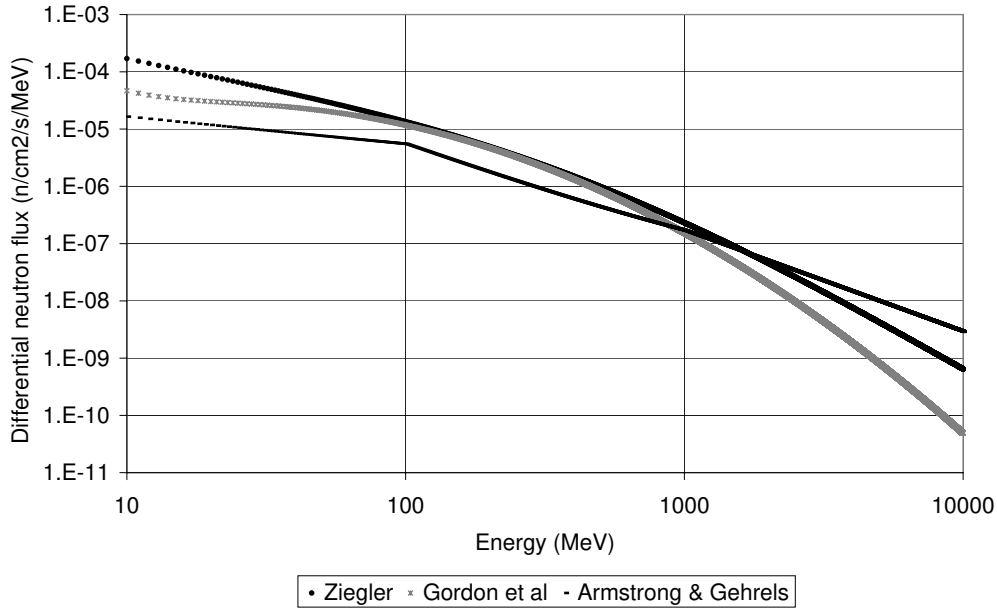}
 \caption{Differential flux of cosmic neutrons at sea level, using the parameterizations from Armstrong and Gehrels \cite{armstrong,gehrels}, Ziegler \cite{ziegler} and Gordon et al.
 \cite{gordon}.}
 \label{spc}
\end{figure}

Different forms of the neutron spectrum at sea level have been used
in cosmogenic activation studies, like the Hess \cite{hess} and
Lal\&Peters \cite{lal} spectra. The ACTIVIA code uses the energy
spectrum from parameterizations in Refs.~\citen{armstrong,gehrels}.
In Ref.~\citen{ziegler}, a compilation of measurements was made,
including the historical Hess spectrum and relevant corrections, and
an analytic function valid from 10 MeV to 10 GeV was derived. In
Ref.~\citen{gordon} a set of measurements of cosmic neutrons on the
ground across the USA was accomplished using Bonner sphere
spectrometers; a different analytic expression fitting data for
energies above 0.4~MeV was proposed (referred hereafter as Gordon et
al. spectrum). Just to appreciate the difference between different
spectra, three considered parameterizations are compared in
Fig.~\ref{spc}, applied for the conditions of New York City at sea
level; the integral flux from 10~MeV to 10~GeV is 5.6$\times
10^{-3}$cm$^{-2}$s$^{-1}$ for Ref.~\citen{ziegler} and 3.6$\times
10^{-3}$cm$^{-2}$s$^{-1}$ for Ref.~\citen{gordon}. The Gordon et al.
parametrization gives a lower neutron flux above 1000~MeV. To
evaluate the activation at a particular location, the relative
neutron flux to the sea-level flux in New York City can be
determined using the calculator available in
Ref.~\citen{calculator}.

For surface protons, the energy spectrum from Ref.~\citen{hagmann}
can be used; the total flux corresponding to the energy range from
100~MeV to 100~GeV is 1.358$\times 10^{-4}$cm$^{-2}$s$^{-1}$. It is
worth noting that in Ref.~\citen{hagmann2}, the energy spectra at
sea level for different particles including neutrons, protons and
muons have been generated from Monte Carlo simulation of primary
protons implementing a model of the Earth's atmosphere in different
codes, finding a satisfactory agreement between codes and available data. Therefore, it seems that Monte Carlo calculations
could provide too quite reliable predictions of cosmic-ray
distributions at sea level.

\section{Germanium}
\label{secge}

A great deal of experiments and projects looking for rare events
(like the interaction of dark matter or the double beta decay of
$^{76}$Ge) have used or are using germanium crystals either as
conventional diodes (like IGEX \cite{igex}, Heidelberg-Moscow
\cite{hm}, GERDA \cite{gerda}, {\sc{Majorana}} \cite{majorana},
CoGENT \cite{cogent}, TEXONO or CDEX \cite{cdex}) or as cryogenic
detectors (like CDMS \cite{supercdms} or EDELWEISS
\cite{edelweiss}). Therefore, many different activation studies have
been performed for germanium. The observation for the first time of
several cosmogenic isotopes through low energy peaks produced due to
their decay by electron capture in the crystals of the CoGENT
experiment was reported in Ref.~\citen{cogent} and cosmogenic
products are usually considered in the background models of
experiments \cite{gerdabkg}. In this section, the main results
obtained for quantifying the cosmogenic yields in both natural and
enriched\footnote{Typical isotopic composition of enriched germanium
used in double beta decay searches is 86\% of $^{76}$Ge and 14\% of
$^{74}$Ge.} germanium are summarized.

\begin{itemlist}
\item Early estimates of production rates of induced isotopes were made in Refs.~\citen{avignone,miley}
using excitation functions calculated with the spallation reaction
code LAHET/ISABEL and the Hess and Lal\&Peters neutron spectra.
Production rates were also derived experimentally in Homestake and
Canfranc laboratories from germanium detectors previously exposed
\cite{avignone}. Agreement with calculation was within a factor of 2
and in some cases within 30\%.

\item Productions of cosmogenic isotopes was assessed in
Ref.~\citen{genius} using a semiempirical code named $\Sigma$.

\item Production cross sections were measured irradiating at Los Alamos Neutron Science Center
(LANSCE) a natural germanium target with a proton beam with an
energy of 800~MeV \cite{norman}. Screening with germanium detectors
was performed at Berkeley intermittently from 2~weeks to 5~years
after irradiation. A quite good agreement with predictions of
Silberberg\&Tsao formulae was obtained.

\item Estimates of production rates for $^{60}$Co and $^{68}$Ge were made using excitation functions calculated with the SHIELD
code in Ref.~\citen{barabanov}. It is worth noting than in this
estimate rates were evaluated including not only the neutron but
also the proton contribution; as mentioned in section~\ref{cal}, the
latter amounts around a $\sim$10\% of the total.

\item The ACTIVIA code, including the energy spectrum for cosmic ray neutrons at sea level based on the
parametrization from Armstrong \cite{armstrong} and Gehrels
\cite{gehrels} was used to evaluate production rates in both natural
and enriched germanium for benchmark \cite{activia}.

\item Another estimate of relevant production rates can be found in
Ref.~\citen{mei}. Excitation functions were calculated using the
TALYS code and the neutron spectrum was considered from the Gordon
et al. parametrization.

\item In Ref.~\citen{elliot2010}, a 11-g sample of enriched germanium was exposed at LANSCE to a
wide-band pulsed neutron beam that resembles the cosmic-ray neutron
flux, with energies up to about 700~MeV. After cooling, germanium
gamma counting was performed underground at the Waste Isolation
Pilot Plant (WIPP) for 49~days to evaluate the nuclei production.
This production was also predicted by calculating cross sections
with CEM03 code. Calculations seem to overestimate production in a
factor of 3 depending on isotope. In addition, measured yields were
converted to cosmogenic production rates considering the Gordon et
al. neutron spectrum.

\item An estimate of production rates after a careful evaluation of excitation
functions was presented in Ref.~\citen{cebrian}. Information on
excitation functions for each relevant isotope was collected
searching for experimental data (available only for protons) and
available calculations (MENDL libraries \cite{mendl} and other ones
based on different codes \cite{study}) and performing some new
calculations (using YIELDX). Then, deviation factors were evaluated
between measured cross sections and different calculations and the
selected description of the excitation functions was the following:
HMS-ALICE calculations for neutrons below 150~MeV and YIELDX results
above this energy. Production rates were computed for both the
Ziegler and Gordon et al. spectra. It was checked that, in general,
estimates based on the Gordon et al. spectrum are closer to
available experimental results; the uncertainty in the activation
yields coming from the cosmic ray flux was for most of the analyzed
products below a factor 2.

\item Neutron irradiation experiments have been performed on
enriched germanium to determine the neutron radiative-capture cross
sections on $^{74}$Ge and $^{76}$Ge, for neutron thermal energies
\cite{ncapture1,ncapture2} and energies of the order of a few MeV
\cite{ncapture3}. These results are of special interest for double
beta decay experiments like GERDA and {\sc{Majorana}}. Neutron
capture results in a cascade of prompt de-excitation gamma rays from
excited states of produced nucleus and the emissions from the decay
of this nucleus, if it is radioactive.

\item Recently, a precise quantification of cosmogenic products in natural germanium including tritium has been made by the EDELWEISS III direct dark matter search
experiment \cite{edelweisscos}, following a detailed analysis of a
long measurement with many germanium detectors with different
exposures to cosmic rays. The cosmogenic yields are evaluated by
fitting the measured data in the low energy region to a continuum
level plus several peaks at the binding energies of electrons in K
and L shells, corresponding to different isotopes decaying by
electron capture. The decay rates measured in detectors are
converted into production rates, considering their well-known
exposure history above ground during different steps of production
and shipment. Estimates of the production rates calculated with the
ACTIVIA code (modified to consider the Gordon et al. cosmic neutron
spectrum) are also presented in Ref.~\citen{edelweisscos} for
comparison.

\item Presence of tritium in the germanium detectors of the {\sc{Majorana Demonstrator}} focused on the study
of the double beta decay has been reported too~\cite{majorana}. A
first estimate of the production rate of several radioisotopes in
enriched germanium, including tritium, from the data of the
{\sc{Majorana Demonstrator}} has been very recently
presented~\cite{majoranalrt}. These enriched detectors (having 87\%
of $^{76}$Ge and 13\% of $^{74}$Ge) have a very well-known exposure
history. The fitting model to derive the abundance of cosmogenic
products is comprised of a calculated tritium beta-decay spectrum,
flat background, and multiple X-ray peaks.

\item A detailed study of the impact of cosmogenic activation in natural and enriched germanium has been carried out in Ref.~\citen{wei}, estimating the production rates of many
isotopes including tritium not only for dominating neutrons but also
for protons and muons, using Geant4 simulations and ACTIVIA
calculations. In addition, the effect of a shielding against
activation has been analyzed and the expected background rates in
the regions of interest have been evaluated for particular exposure
and cooling conditions. All results from this work considered here
correspond to neutron activation using the Gordon et al. spectrum.

\end{itemlist}

Tables~\ref{genat} and~\ref{geenr} summarize the rates of production
(expressed in kg$^{-1}$ d$^{-1}$) of some of the isotopes induced in
natural and enriched germanium at sea level obtained either from
measurements or from calculations based on the different approaches
described before. Experimental results quoted in
Ref.~\citen{heusser} are also included. When comparing all the
available results for germanium, the order of magnitude of measured
values is reasonably reproduced by calculations, but there is an
important dispersion in results for some isotopes which can be very
relevant, like $^{68}$Ge. In enriched germanium, cosmogenic
activation is significantly suppressed for most of the isotopes.

\begin{table}
\tbl{Production rates (in kg$^{-1}$ d$^{-1}$) at sea level for
isotopes induced in natural germanium following measurements in
Refs.~\citen{avignone,edelweisscos} and different calculations (see
text). } {\begin{tabular}{@{}lcccccccc@{}} \toprule &
$^{49}$V & $^{54}$Mn & $^{55}$Fe & $^{57}$Co & $^{58}$Co & $^{60}$Co & $^{65}$Zn & $^{68}$Ge \\
\colrule Half-life\cite{toi,ddep} & 330~d & 312.19~d & 2.747~y &
271.82~d & 70.85~d & 5.2711~y & 244.01~d & 270.95~d  \\
Measurement \cite{avignone} & &  3.3$\pm$0.8 & &  2.9$\pm$0.4 &  3.5$\pm$0.9 & & 38$\pm$6 & 30$\pm$7 \\
Measurement \cite{edelweisscos} & 2.8$\pm$0.6 & & 4.6$\pm$0.7 & & & & 106$\pm$13 & $>$74 \\
Monte Carlo \cite{avignone} &  & 2.7 & & 4.4 &   5.3  &  &  34.4 &  29.6       \\
Monte Carlo \cite{miley} &   & & & 0.5 & 4.4 &  4.8 & 30.0 & 26.5 \\
Sigma \cite{genius} &   & 9.1 & 8.4 & 10.2 & 16.1 & 6.6 &79.0 & 58.4
\\ 
SHIELD \cite{barabanov} &  & &  & & & 2.9  &  & 81.6  \\
ACTIVIA \cite{activia} & & 2.7 & 3.4 & 6.7 & 8.5 & 2.8 & 29.0 & 45.8 \\
TALYS \cite{mei} &  & 2.7  & 8.6  & 13.5   & & 2.0 & 37.1  & 41.3 \\
MENDL+YIELDX \cite{cebrian} &  & 5.2 &   6.0 & 7.6 & 10.9   & 3.9  & 63  & 60 \\
ACTIVIA \cite{edelweisscos} & 1.9 & & 3.5 & & & & 38.7 & 23.1  \\
ACTIVIA (MENDL-2P) \cite{edelweisscos} & 1.9 & & 4.0 & & & & 65.8 & 45.0 \\
GEANT4 \cite{wei} & & 2.0 & 7.9 & 7.4 & 5.7 & 2.9 & 75.9 & 182.8 \\
ACTIVIA \cite{wei} & & 2.8 & 4.1 & 8.9 & 11.4 & 4.1 & 44.2 & 24.7 \\

 \botrule
\end{tabular} \label{genat}}
\end{table}

\begin{table}
\tbl{Production rates (in kg$^{-1}$ d$^{-1}$) at sea level for
isotopes induced in enriched germanium (86-87\% of $^{76}$Ge and
14-13\% of $^{74}$Ge) following measurements quoted in
Refs.~\citen{heusser,elliot2010,majoranalrt} and different
calculations (see text). } {\begin{tabular}{@{}lccccccc@{}} \toprule
&  $^{54}$Mn & $^{55}$Fe & $^{57}$Co & $^{58}$Co & $^{60}$Co & $^{65}$Zn & $^{68}$Ge \\
\colrule Half-life\cite{toi,ddep} & 312.19~d & 2.747~y & 271.82~d &
70.85~d & 5.2711~y & 244.01~d & 270.95~d  \\
Measurement \cite{heusser} & 2.3 & &1.6 & 1.2 & & 11 & \\
Measurement \cite{elliot2010} & 2.0$\pm$1.0 & & 0.7$\pm$0.4 &  & 2.5$\pm$1.2 & 8.9$\pm$2.5 & 2.1$\pm$0.4 \\
Measurement \cite{majoranalrt} & 4.4$\pm$4.1 & 2.1$\pm$0.7 & & & & 4.3$\pm$3.6 & 3.3$\pm$1.6 \\
Monte Carlo \cite{avignone} & 1.4 && 1 &1.8 &   &  6.4 &0.94  \\
Monte Carlo \cite{miley} && & 0.08 & 1.6 &  3.5 & 6.0 & 1.2 \\
SHIELD \cite{barabanov}  &&  &  && 3.3 & & 5.8 \\
ACTIVIA \cite{activia} & 2.2 & 1.6 & 2.9 & 5.5 & 2.4 & 10.4 & 7.6 \\
TALYS \cite{mei} & 0.87 & 3.4 & 6.7 &  & 1.6  & 20 & 7.2 \\
MENDL+YIELDX \cite{cebrian} & 3.7 & 1.6 & 1.7 & 4.6 & 5.1 & 20 & 12 \\
GEANT4 \cite{wei} & 1.4 & 4.5 & 3.3 & 2.9 & 2.4 & 24.9 & 21.8 \\
ACTIVIA \cite{wei} & 2.2 & 1.2 & 2.3 & 5.5 & 4.4 & 9.7 & 15.4 \\
 \botrule
\end{tabular} \label{geenr}}
\end{table}

As pointed out before, germanium is being used as a target for dark
matter searches for many years, either as pure ionization detectors
or in cryogenic detectors measuring simultaneously ionization and
heat. The production of tritium in the detector media can be
specially worrisome due to its emissions. A part of the unexplained
background in the low energy region of the IGEX detectors could be
attributed to tritium \cite{cebriantaup2003} and tritium is
highlighted as one the relevant background sources in future
experiments like SuperCDMS \cite{supercdms}. Indeed, the first
experimental estimates of the tritium production rate in germanium
at sea level have been derived by the EDELWEISS collaboration
\cite{edelweisscos} for natural germanium and using data from the
{\sc{Majorana Demonstrator}} \cite{majoranalrt} for enriched
germanium; a new estimate based on CDMSlite data \cite{cdmslite} has been also presented. Several calculations had been made before: a rough
calculation was attempted in Ref.~\citen{avignone} using two
different neutron spectra. This was followed by other estimates
based in TALYS \cite{mei} and ACTIVIA and GEANT4 \cite{mei2016,wei},
profiting from the availability of reaction codes that fully
identify all the reaction products in the final state. In
Ref.~\citen{tritiumpaper}, dedicated calculations of production
rates of tritium at sea level have been performed for some of the
materials typically used as targets in dark matter detectors
(germanium, sodium iodide, argon and neon), based on a selection of
excitation functions over the entire energy range of cosmic
neutrons, mainly from TENDL for neutrons and HEAD-2009 libraries.
All these results on the production rate of tritium in both natural
and enriched germanium are summarized in Table~\ref{tritium}. An
additional estimate of the production rate in enriched germanium
using the COSMO code gives 70~kg$^{-1}$ d$^{-1}$
\cite{cebriantaup2003}. For the natural material, calculations give in general lower values than the measured rates; in
particular, the smallest value from \cite{mei} can be understood
because using TALYS cross-sections only contributions from the
lowest energy neutrons are considered. The range derived in
Ref.~\citen{tritiumpaper} is well compatible with the measured rates
by EDELWEISS and CDMSlite. For enriched germanium, as used in double beta decay
experiments, the measured production rate is higher than in natural
germanium; this can be due to the fact that cross sections increase
with the mass number of the germanium isotope in all the energy
range above $\sim$50~MeV, according to TENDL-2013 and HEAD-2009 data
\cite {tritiumpaper}.

\begin{table}
\tbl{Production rates (in kg$^{-1}$d$^{-1}$) of $^{3}$H at sea level
evaluated for several targets from measurements or calculations
based on different approaches (see text). The two values from
Ref.~\citen{avignone} were derived using different neutron spectra
and the two values from ACTIVIA in Ref.~\citen{edelweisscos}
correspond to using just semiempirical cross sections or data from
MENDL-2P too.}
{\begin{tabular}{@{}lccccccccc@{}} \toprule Target  & Ref.~\citen{avignone} & TENDL+HEAD\cite{tritiumpaper} & TALYS\cite{mei}  & GEANT4\cite{mei2016} & GEANT4\cite{wei} & ACTIVIA\cite{mei2016}  & ACTIVIA\cite{wei} & ACTIVIA & Others  \\
\colrule $^{nat}$Ge & 178/210 &75$\pm$26  & 27.7 & 48.3 & 47.4 &
52.4 & 52.4 & 46/43.5 \cite{edelweisscos} & 82$\pm$21 \cite{edelweisscos}\\
& & & & & & & & &  76$\pm$6 \cite{cdmslite}\\
$^{enr}$Ge & 113/140 & 94$\pm$34 & 24.0 & & 47.4 & & 51.3 & & 140$\pm$10 \cite{majoranalrt} \\
Si && & & 27.3 && 108.7 && & 125 \cite{supercdms} \\
TeO$_{2}$ && & 43.7 & & & &&& \\
NaI && 83$\pm$27  & 31.1 & 42.9 &&  36.2 && 26 \cite{pettus} &\\
CsI && & 19.7 & & & & &&\\
CaWO$_{4}$ && & 45.5 & &&& & & \\
Ar && 146$\pm$31 & 44.4 & 84.9 && 82.9 &&& \\
Ne && 228$\pm$16 & & & &&&& \\
Xe && & 16.0 & 31.6 && 35.6 && & \\
Quartz && & & & &&& 46 \cite{pettus} & \\
C$_{2}$H$_{6}$ && & & 279.5 &&& & & \\ \botrule
\end{tabular} \label{tritium}}
\end{table}

In addition to the quantification of the cosmogenic activation, the
effect of this activity in germanium detectors has been studied too
\cite{wei}. In Ref.~\citen{jian}, a background simulation method for
cosmogenic radionuclides inside HPGe detectors for rare event
experiments is presented, to quantify the expected spectrum for each
internal cosmogenic isotope analyzing phenomena caused by the
coincidence summing-up effect.

\section{Silicon}
\label{secsi}

Silicon is the detector medium in cryogenic detectors, like those of
the CDMS experiment, or in CCD detectors, used in the DAMIC
experiment \cite{damic2}, both devoted to the direct detection of
dark matter. The intrinsic isotope $^{32}$Si, a $\beta^{-}$ emitter
with an endpoint energy of 224.5~keV and half-life of 150~y
\cite{toi}, can be a relevant background source. It is produced as a
spallation product from cosmic ray secondaries on argon in the
atmosphere; $^{32}$Si atoms can make their way into the terrestrial
environment through aqueous transport and therefore its
concentration can depend on the exact source and location of the
silicon used in the production and fabrication of detectors. A decay
rate of 80$_{-65}^{+110}$~kg$^{-1}$ d$^{-1}$ has been found by the
DAMIC collaboration in their CCD detectors \cite{damic} and this
isotope is considered in the estimates of SuperCDMS sensitivity
\cite{supercdms}.

Also tritium can be relevant for silicon detectors, as pointed out
in Ref.~\citen{supercdms}. An estimate of the production rate based
on GEANT4 and ACTIVIA (using the Gordon et al. neutron spectrum) is
presented in Ref.~\citen{mei2016} and results are shown in
Table~\ref{tritium}, together with the production rate considered by
the SuperCDMS collaboration \cite{supercdms}.

\section{Tellurium}
\label{secte}

Activation in tellurium has been mainly studied with focus on the
CUORE \cite{cuore} and SNO+ \cite{sno} double beta decay
experiments. The production rate of tritium in TeO$_{2}$ was also
evaluated using the TALYS code \cite{mei} and it is reported in
Table~\ref{tritium}.

Results for proton production cross sections obtained within the
CUORE project were published in Ref.~\citen{te}, completing previous
results on proton spallation reactions on tellurium
\cite{bardayan,norman}. Some measurements were made in USA, by
irradiating a tellurium target with the 800~MeV proton beam at
LANSCE and performing a gamma screening with germanium detectors in
Berkeley. Other measurements were carried out in Europe, exposing
TeO$_{2}$ targets to proton beams with energies of 1.4 and 23~GeV at
CERN; first germanium screening was made there for several months
and later in Milano 2.8 and 4.6~years after irradiation. The
obtained cross sections at the three energies are in good agreement
with Silberberg\&Tsao predictions. 
In addition, as presented in Ref.~\citen{ten}, flux-averaged
cross-sections for neutron activation in natural tellurium were
measured by irradiating at LANSCE TeO$_{2}$ powder with the neutron
beam containing neutrons of kinetic energies up to $\sim$800 MeV and
having an energy spectrum similar to that of cosmic-ray neutrons at
sea-level; gamma ray analysis was performed in Berkeley Low
Background Facility. The cross sections obtained for $^{110m}$Ag and
$^{60}$Co, the two isotopes identified as the most relevant ones in the background model of the CUORE experiment
\cite{cuorebkg}, were combined with results from tellurium
activation measurements with 800~MeV-23~GeV protons to estimate the
corresponding production rates for CUORE crystals, considering the
Gordon et al. spectrum. These results are reported in
Table~\ref{te}.

Within the SNO+ project, as described in Ref.~\citen{telozza},
production rates of cosmogenic-induced isotopes on a natural
tellurium target were calculated using the ACTIVIA program for
energies above 100~MeV, the neutron and proton cross sections from
TENDL library in the 10-200~MeV range when available, and the
cosmogenic neutron and proton flux parametrization at sea level from
Armstrong \cite{armstrong} and Gehrels \cite{gehrels}. An extensive
set of isotopes was considered and even expected activation
underground was evaluated too. Some results are shown in
Table~\ref{te}. Although not directly comparable with the estimated
rates in Ref.~\citen{ten} for the different targets, there is an
important discrepancy between them. As shown in the excitation
functions plotted in Ref.~\citen{telozza} for $^{110m}$Ag and
$^{60}$Co, the measured cross sections above 800~MeV agree
reasonably well with predictions of ACTIVIA; the origin of the
difference can be at lower energies (the region giving the dominant
contribution to the production rates). The different cosmic neutron
spectra considered in the two works makes difficult also the
comparison.

\begin{table}
\tbl{Production rates (in $\mu$Bq kg$^{-1}$) at sea level presented
in Ref.~\citen{ten} for isotopes induced in TeO$_{2}$ combining
measured cross sections at different energies and in
Ref.~\citen{telozza} for natural tellurium using ACTIVIA and TENDL
library. Different cosmic neutron spectra were considered in the two
works.} {\begin{tabular}{@{}cccc@{}}
\toprule Isotope & Half-life\cite{ddep} & Ref.~\citen{ten} & Ref.~\citen{telozza} \\
\colrule
$^{110m}$Ag & 249.78~d & 4.9 10$^{-3}$ & 2.39 \\
$^{60}$Co & 5.2711~y  & $<$6.3 10$^{-5}$  & 0.81 \\ \botrule 
\end{tabular} \label{te}}
\end{table}

\section{Sodium Iodide}
\label{secnai}

Sodium iodide is being used for dark matter searches for a long time
and in the last years activation yields in these crystals have
started to be quantified in the context of various experiments. The
observation of the annual modulation signal registered in the data
of the DAMA/LIBRA experiment \cite{damaresults} for many years is
the goal of projects at different stages of development like KIMS
\cite{kims} and DM-Ice \cite{dmice} (now joint in COSINE), ANAIS
\cite{anais}, SABRE \cite{sabre}, PICO-LON \cite{picolon} and
COSINUS \cite{cosinus}.

A first direct estimate of production rates of several iodine,
tellurium and sodium isotopes induced in NaI(Tl) crystals was
presented in Ref.~\citen{cebrianjcap}. The estimate, developed in
the frame of the dark matter ANAIS experiment, was based on data
from two 12.5~kg NaI(Tl) detectors produced by Alpha Spectra Inc.
which were installed inside a convenient shielding at the Canfranc
Underground Laboratory at the end of 2012, just after finishing
surface exposure to cosmic rays during production. The very fast
start of data taking allowed to identify and quantify isotopes with
half-lives of the order of tens of days. Initial activities
underground were measured following the evolution of the identifying
signatures for each isotope along several months; then, production
rates at sea level were properly estimated according to the history
of detectors. Production rates were also computed for comparison
using the cosmic neutron flux at sea level from Gordon et al. and a
description of excitation functions, carefully selected minimizing
deviation factors between measured cross sections and calculations
from MENDL2N, TENDL-2013, YIELDX and HEAD-2009. The ratio between
calculated and measured production rates ranges from 0.7 to 1.5 for
the observed isotopes. The obtained rates are reported in
Table~\ref{nai}.

A complete study of cosmogenic activation in NaI(Tl) crystals was
also performed within the DM-Ice17 experiment \cite{pettus}. Two
crystals, with a mass of 8.5~kg each and previously operated at the
dark matter NaIAD experiment in the Boulby underground laboratory in
UK \cite{naiad}, were deployed at a depth of 2450~m under the ice at
the geographic South Pole in December 2010 and collected data over
three-and-a-half years. The activation of detector components
occurred during construction, transportation, or storage at the
South Pole prior to deployment. The DM-Ice17 data provided
compelling evidence of significant production of cosmogenic
isotopes, which was used to derive the corresponding production
rates based on simulation matching. In addition, calculations based
on modified ACTIVIA using the Gordon et al. neutron spectrum were
also carried out. Results are presented in Table~\ref{nai}. A very
good agreement between measured production rates in ANAIS and
DM-Ice17 crystals has been found. ACTIVIA estimates of metastable
tellurium isotopes give rates clearly higher than measured values.

\begin{table}
\tbl{Production rates (in kg$^{-1}$d$^{-1}$) at sea level for
isotopes induced in NaI(Tl) crystals following measurements by ANAIS \cite{cebrianjcap}
and DM-Ice17 \cite{pettus} experiments together with calculations based on
selected excitation functions \cite{cebrianjcap} or using the
ACTIVIA code \cite{pettus}. } {\begin{tabular}{@{}lccccc@{}}
\toprule Isotope & Half-life\cite{toi,ddep} &
Calculation\cite{cebrianjcap} & Measurement\cite{cebrianjcap} &
ACTIVIA\cite{pettus} &  Measurement\cite{pettus}\\
\colrule

$^{126}$I & 12.93~d &  297.0 & 283$\pm$36  & 128 &   \\

$^{125}$I & 59.407~d &  242.3  & 220$\pm$10 & 221 & 230  \\


$^{127m}$Te & 107~d &  7.1 &  10.2$\pm$0.4 & 93 & $<$9\\

$^{125m}$Te & 57.4~d &   41.9 &  28.2$\pm$1.3 & 74 & 27 \\

$^{123m}$Te & 119.3~d &  33.2  &  31.6$\pm$1.1 & 52 & 21 \\


$^{121m}$Te & 154~d &  23.8  &  23.5$\pm$0.8 & 93 & 25 \\

$^{121}$Te & 19.16~d &   8.4  & 9.9$\pm$3.7 & 93 &  \\

$^{113}$Sn & 115.09~d &  & & 9.0 & 16 \\

$^{22}$Na & 2.6029~y &  53.6 & 45.1$\pm$1.9 & 66 & \\

\botrule
\end{tabular} \label{nai}}
\end{table}

Presence of cosmogenic isotopes has been also observed in NaI(Tl)
crystals from other experiments
\cite{kims,kims2,damanima2008,damaijmpa}. In particular, in
Ref.~\citen{damanima2008}, the fraction of $^{129}$I (which can be
produced by uranium spontaneous fission and by cosmic rays) in
DAMA/LIBRA crystals from Saint Gobain company was determined to be
$^{129}I/^{nat}I=(1.7\pm0.1)\times 10^{-13}$; strong variability of
this concentration is expected in different origin ores.

The tritium content in NaI(Tl) detectors devoted to dark matter
searches could be very relevant. The tritium activity in DAMA/LIBRA
crystals was constrained to be $<$0.09~mBq/kg (95\% C.L.)
\cite{damanima2008}. A direct identification of tritium in NaI(Tl)
crystals has not been possible, but within the ANAIS experiment, the
construction of a detailed background model of the operated modules
in the Canfranc Underground Laboratory (based on a Geant4 simulation
of quantified background components) points to the need of an
additional background source contributing only in the very low
energy region, which could be tritium \cite{anaisepjc}. The
inclusion of a certain activity of $^{3}$H homogeneously distributed
in the NaI crystal provides a very good agreement between measured
data and the model below 20~keV. For two Alpha Spectra detectors,
both fabricated in Alpha Spectra facilities at Grand Junction,
Colorado (where the cosmic neutron flux is estimated to be a factor
$f=3.6$ times higher than at sea level \cite{cebrianjcap}) but with
different production history, the required $^{3}$H initial
activities (that is, at the moment of going underground) to
reproduce the data are around 0.20~mBq/kg and 0.09~mBq/kg; the
latter value is just the upper limit set for DAMA/LIBRA crystals.
These results boosted the study of tritium production in NaI.

Only calculations of the tritium production rate are available;
results using TALYS \cite{mei}, GEANT4 \cite{mei2016} and ACTIVIA
\cite{mei2016,pettus} are shown in Table~\ref{tritium}. In the
calculations of Ref.~\citen{tritiumpaper}, based on a selection of
excitation functions mainly from TENDL for neutrons and HEAD-2009
libraries, there is a significant difference in the medium energy
range between the two estimates of cross sections for I, being those
from TENDL much higher; this fact is responsible of the large
uncertainty in the production rate estimated for NaI. As it can be
seen in Table~\ref{tritium}, this rate is higher than previous
estimates, but the required exposure times to get the deduced
tritium activities for ANAIS detectors based on that rate roughly
agree with the time lapse between sodium iodide raw material
purification starting and detector shipment, according to the
company \cite{tritiumpaper}.

\section{Xenon}
\label{secxe}

Xenon-based detectors are being extensively used in the
investigation of rare events. Following very successful
predecessors, many efforts are now concentrated in the XENON1T
\cite{xenon1t}, LUX-ZEPLIN \cite{lztdr}, PANDAX \cite{pandax} or
XMASS \cite{xmass} projects in the search for dark matter; EXO
\cite{exo}, KamLAND-Zen \cite{kamlandzen} and NEXT \cite{next} are
looking for the neutrinoless double beta decay of $^{136}$Xe. Even
if in liquid xenon experiments, the purification system is presumed
to suppress the concentration of non-noble radioisotopes below
significance, there are several estimates of production rates of
activated isotopes in xenon based on both direct measurements and
calculations following different approaches. As it will be shown,
the dispersion in results for most of the analyzed products in this
material is important. Production of xenon radioisotopes could be
problematic due to low energy deposits, which can be a relevant
background in the WIMP search energy region; but these backgrounds
soon reach negligible levels once xenon is moved underground
\cite{lztdr}.

A first dedicated study on the cosmogenic activation of xenon was
presented in Ref.~\citen{baudis}, complemented by a study of copper
activated simultaneously. Samples were exposed to cosmic rays in a
controlled manner at the Jungfraujoch research station (at 3470~m
above sea level in Switzerland) for 345~days. The xenon sample, with
a mass of $\sim$2~kg, had natural isotopic composition. The samples
were screened with a low-background germanium detector (named Gator)
in the Gran Sasso underground laboratory before and after activation
to quantify the cosmogenic products. Saturation activities were
derived for several long-lived isotopes and some results are shown
in Table~\ref{xe}. The measured results were compared in addition to
predictions using ACTIVIA and COSMO packages, including the neutron
spectrum from Refs.~\citen{armstrong,gehrels}. It is worth noting
that from all the directly observed radionuclides, only $^{125}$Sb
is considered to be a potential relevant background for a multi-ton
scale dark matter search.

For natural xenon, production rates of several isotopes were
estimated as made for germanium in Ref.~\citen{mei}, with excitation
functions calculated using TALYS code, and as for different
materials in Ref.~\citen{mei2016}, from GEANT4 simulation or ACTIVIA
calculations. For GEANT4 simulations, the set of electromagnetic and
hadronic physics processes included in the Shielding modular physics
list were taken into account while in ACTIVIA, cross sections are
obtained from data tables and semiempirical formulae. It is worth
noting that in the results shown here the Gordon et al. neutron
spectrum was considered for both GEANT4 and ACTIVIA, which required
a modification of the latter. Some of these results are also shown
in Table~\ref{xe}. In these calculations, the production rate of
$^{3}$H was also evaluated and it is presented in
Table~\ref{tritium}.

The analysis of the radiogenic and muon-induced backgrounds in the
LUX dark matter detector operating at the Sanford Underground
Research Facility in US allowed to quantify the cosmogenic
production of some xenon radioisotopes \cite{lux}. In particular,
zero-field data taken 12~days after the xenon was moved underground
and 70 days before the start of the WIMP search run were considered.
Some of the obtained saturation activities, as reproduced in
Ref.~\citen{baudis}, are presented in Table~\ref{xe}.

\begin{table}
\tbl{Production rates or saturation activities (in $\mu$Bq
kg$^{-1}$) at sea level for isotopes induced in natural xenon
following measurements in Refs.~\citen{baudis,lux} and different
calculations based on COSMO, ACTIVIA (considering different cosmic
neutron spectra), TALYS and GEANT4. }
{\begin{tabular}{@{}lcccccc@{}} \toprule  &
$^{7}$Be & $^{101}$Rh & $^{125}$Sb & $^{121m}$Te & $^{123m}$Te & $^{127}$Xe \\
\colrule

Half-life\cite{toi,ddep} & 53.22~d & 3.3~y & 2.759~y & 154~d & 119.3~d & 36.358~d  \\

Measurement \cite{baudis} & 370$^{+240}_{-230}$ & 1420$^{+970}_{-850}$ & 590$^{+260}_{-230}$ & $<$1200 & $<$610 & 1870$^{+290}_{-270}$  \\

Measurement \cite{lux} & & & & & &1530$\pm$300  \\

COSMO \cite{baudis} & 6.4 & 15.3 & 13.5 & 276 & 14.4 & 555  \\

ACTIVIA \cite{baudis} & 6.4 & 16.6 & 0.2 & 299 & 14.7 & 415  \\

ACTIVIA \cite{mei2016} & & & 0.10 & 630.3 & 30.9 & 1041.0 \\

GEANT4 \cite{mei2016} & & & 17.1 & 245.4 & 213.8 & 2700.2 \\

TALYS \cite{mei} & & 0.46 & 0.46 & 135.4 & 140.0 & \\
\botrule
\end{tabular} \label{xe}}
\end{table}

In the context of neutrinoless double beta decay experiments working
with xenon enriched in $^{136}$Xe, the production of short-lived
$^{137}$Xe by neutron capture can be relevant and is considered as a
background source. This isotope is the only cosmogenic radionuclide
found to have a significant contribution in their region of interest
\cite{exobkg}; the measured capture rate by the EXO-200 experiment
is 338$^{+132}_{-93}$ captures on $^{136}$Xe per year. In addition,
individual production cross sections for 271 radionuclides have been
determined for $^{136}$Xe in an inverse kinematics experiment at
GSI\cite{xe136}.

\section{Argon}
\label{secar}

Different projects use liquid argon in dark matter detectors, like
ArDM \cite{ardm}, DarkSide \cite{darkside2} or DEAP/CLEAN
\cite{deap}, and also gaseous argon is considered in the TREX-DM
experiment \cite{trexdm}. The most worrisome background source in
this detector medium is $^{39}$Ar, a $\beta^{-}$ emitter with an
endpoint energy of 565~keV and half-life of 269~y \cite{toi}.
$^{39}$Ar is present in atmospheric argon as it is mainly produced
by cosmic ray induced nuclear reactions such as
$^{40}$Ar(n,2n)$^{39}$Ar. In the context of the WARP experiment at
the Laboratori Nazionali del Gran Sasso (LNGS), $^{39}$Ar activity
in natural argon was measured to be at the level of 1~Bq/kg
\cite{warp}. But the discovery of argon from deep underground
sources with significantly less $^{39}$Ar content has allowed to
improve the background prospects of experiments using argon as the
active target. In Ref.~\citen{darkside1}, a specific $^{39}$Ar
activity of less than 0.65\% of the activity in atmospheric argon
corresponding to 6.6~mBq/kg was measured for underground argon from
a CO$_{2}$ plant in Colorado; results from the DarkSide experiment
have shown that the underground argon contains $^{39}$Ar at a level
reduced by a factor (1.4$\pm$0.2$)\times$10$^{3}$ relative to
atmospheric one \cite{darkside2}.

Concerning the tritium production in argon (or neon, which is in
some cases considered as an alternative target), there is no
experimental information. As shown in Table~\ref{tritium}, there are
some estimates of the production rate in argon using TALYS code
\cite{mei} and GEANT4 and ACTIVIA (using the Gordon et al. neutron
spectrum) \cite{mei2016}. Since there was no information for neon,
in Ref.~\citen{tritiumpaper}, the study of tritium production
performed for other targets like germanium and sodium iodide was
applied also to argon and neon, selecting the excitation functions
and considering the Gordon et al. cosmic neutron spectrum. If
saturation activity was reached for tritium, according to the
production rates deduced in this study and presented in
Table~\ref{tritium} too, tritium could be very problematic. However,
tritium is expected to be suppressed by purification of gas and
minimizing exposure to cosmic rays of the purified gas should avoid
any problematic tritium activation.

\section{Copper}
\label{seccu}

Copper is a material widely used in rare event experiments due to
its mechanical, thermal and electrical properties, either as shield
or part of the detector components. Copper is also a specially
interesting material for activation studies because there are many
extensive sets of measurements of production cross sections for
protons and even for neutrons; this makes it very attractive to
compare calculations and experimental data in order to allow a good
validation of excitation functions. The most relevant results on the
quantification of activation yields in copper are presented in this
section.

\begin{itemlist}
\item As for germanium, the ACTIVIA code, including the energy
spectrum for cosmic ray neutrons at sea level from
Refs.~\citen{armstrong,gehrels}, was used to evaluate production
rates in copper for benchmark \cite{activia}.

\item Direct measurements of production rates were made in LNGS
\cite{lngsexposure}. Seven plates made of NOSV grade copper from
Nord-deutsche Affinerie AG (Germany), with a total weight of 125~kg,
were exposed for 270~days at an outside hall of the LNGS (altitude
985~m) under a roof. Screening with one GeMPI detector was carried
out for 103~days. Production rates were derived as the measured
saturation activity. The highest values were found for cobalt
isotopes; in particular, the measured activity of $^{60}$Co
((2100$\pm$190)~$\mu$Bq/kg) greatly exceeded the upper limit derived
for the primordial activity. The production rates at sea level are
reproduced in Table~\ref{cu}, including a correction factor
estimated to be 2.1 (following Ref.~\citen{ziegler}) due to the
altitude during exposure.

\item The same study of evaluation of excitation functions based of
deviation factors and estimate of production rates made for
germanium was also carried out for copper in Ref.~\citen{cebrian}.
Production rates were calculated using below 100~MeV MENDL2N results
for neutrons normalized to the available experimental data if
possible, and above that energy experimental data for protons
combined with YIELDX calculations when necessary. Differences in the
production rates estimated in this work due to the different neutron
spectra considered are similar to those obtained in germanium:
production rates considering the Ziegler parametrization are higher
than the corresponding ones using the expression from Gordon et al.
Results shown in Table~\ref{cu} are the ones obtained using the
Gordon spectrum.

\item Together with the dedicated study on the cosmogenic activation
of xenon, a study on copper was simultaneously carried out in
Ref.~\citen{baudis}. Samples were exposed to cosmic rays in a
controlled manner at the Jungfraujoch research station (at 3470~m
above sea level in Switzerland) for 345~days. The 10.35~kg copper
sample consisted of 5 blocks of OFHC copper from Norddeutsche
Affinerie (now Aurubis), from the batch used to construct inner
parts of the XENON100 detector. Before each measurement, copper was
properly cleaned to remove surface contaminations. The samples were
screened with the Gator low-background germanium detector in the
Gran Sasso underground laboratory before and after activation to
quantify the cosmogenic products. Saturation activities were derived
and some results are presented in Table~\ref{cu}. The measured
results were compared in addition to predictions using ACTIVIA and
COSMO packages, including the neutron spectrum from
Refs.~\citen{armstrong,gehrels}.

\item As for different materials, production rates of several isotopes
were estimated in Ref.~\citen{mei} using TALYS and in
Ref.~\citen{mei2016} from GEANT4 simulation or ACTIVIA calculations,
considering the Gordon et al. parametrization for the neutron
spectrum. Some of these results are also shown in Table~\ref{cu}.

\item In Ref.~\citen{ivan}, the cosmogenic activity produced in copper
to be used as radiation shielding was measured, using a HPGe
detector operated in the Canfranc Underground Laboratory. A sample
with a mass of 18~kg, exposed at 250~m above sea level for 1~year
after casting according to company records, as well as bricks
exposed to cosmic rays for 41~days, were analyzed. Activities found
for cobalt isotopes and $^{54}$Mn agree with the expectations from
previously measured production rates.
\end{itemlist}

\begin{table}
\tbl{Production rates (in kg$^{-1}$ d$^{-1}$) at sea level for
isotopes induced in natural copper following measurements in
Refs.~\citen{lngsexposure,baudis} and different calculations (see
text). } {\begin{tabular}{@{}lcccccccc@{}} \toprule &
$^{46}$Sc & $^{48}$V & $^{54}$Mn & $^{59}$Fe & $^{56}$Co & $^{57}$Co & $^{58}$Co & $^{60}$Co \\
\colrule
Half-life\cite{toi,ddep} & 83.787~d & 15.9735~d & 312.19~d & 44.494~d & 77.236~d &  271.82~d & 70.85~d & 5.2711~y   \\
Measurement \cite{lngsexposure} & 2.18$\pm$0.74 & 4.5$\pm$1.6 & 8.85$\pm$0.86 &  18.7$\pm$4.9 & 9.5$\pm$1.2 & 74$\pm$17 & 67.9$\pm$3.7 &  86.4$\pm$7.8 \\
Measurement \cite{baudis} & 2.33$^{+0.95}_{-0.78}$ & 3.4$^{+1.6}_{-1.3}$ & 13.3$^{+3.0}_{-2.9}$ & 4.1$^{+1.4}_{-1.2}$ & 9.3$^{+1.2}_{-1.4}$ & 44.8$^{+8.6}_{-8.2}$ & 68.9$^{+5.4}_{-5.0}$ & 29.4$^{+7.1}_{-5.9}$ \\
ACTIVIA \cite{activia,baudis} & 3.1 & & 14.3 & 4.2 & 8.7 & 32.5 & 56.6 & 26.3 \\
ACTIVIA (MENDL-2P) \cite{activia} & 3.1 & & 12.4 & 1.8 & 14.1 & 36.4 & 38.1 & 9.7 \\
TALYS \cite{mei} & & &  16.2 &  & & 56.2 &  &  46.4 \\
MENDL+YIELDX \cite{cebrian} &  2.7 &  & 27.7 & 4.9 &  20.0 & 74.1 & 123.0 &   55.4 \\
COSMO \cite{baudis} & 1.5 & 3.1 & 13.5 & 4.3 & 7.0 & 30.2 & 54.6 & 25.7 \\
GEANT4 \cite{mei2016} & 1.2 & & 12.3 & 8.8 & 10.3 & 67.2 & 57.3 & 64.6 \\
ACTIVIA \cite{mei2016} & 4.1 & & 30.0 &  10.5 & 20.1 & 77.5 & 138.1 & 66.1 \\
 \botrule
\end{tabular} \label{cu}}
\end{table}

Comparing all the information on production rates on copper
available and summarized in Table~\ref{cu}, it can be seen that
measured rates from the two independent activation experiments are
in very good agreement for some isotopes, but there are differences
for others, specially for $^{60}$Co. In general, higher rates were
found in Ref.~\citen{lngsexposure}. The different calculations give
results with important dispersion; it must be noted that ACTIVIA
calculations reported in Table~\ref{cu} were obtained using
different descriptions of the cosmic neutron spectrum.

\section{Lead}
\label{secpb}

Even if tons of lead are commonly used in rare event experiments as
radiation shielding, activation studies on this material are scarce.
In Ref.~\citen{giuseppe}, results for some production rates are
presented following the irradiation of a natural lead sample at
LANSCE using the neutron beam resembling the cosmic neutron flux and
after counting the amount of radioactive isotopes using a low
background, underground germanium detector at WIPP. By scaling the
LANSCE neutron flux to a cosmic neutron flux, the sea level
production rates of some long-lived radionuclides were estimated and
are reported in Table~\ref{pb}; calculations based on TALYS code and
the Gordon et al. neutron spectrum deduced in the same work are also
shown. In Ref.~\citen{giuseppe} it is concluded that for ordinary
exposures, the cosmogenic background is less than that from the
naturally occurring radioisotopes in lead.

\begin{table}
\tbl{Production rates (in kg$^{-1}$d$^{-1}$) at sea level for
isotopes induced in natural lead following an irradiation
experiment at LANSCE and TALYS calculations.}
{\begin{tabular}{@{}cccc@{}} \toprule Isotope & Half-life (y)
\cite{toi,ddep} & Measurement \cite{giuseppe} & TALYS
\cite{giuseppe}
\\ \colrule
$^{194}$Hg & 444 & 8.0$\pm$1.3 & 16 \\
$^{202}$Pb & 5.25 10$^{4}$  & 120$\pm$25 & 77 \\
$^{207}$Bi & 32.9  & $<$0.17 &  \\ \botrule 
\end{tabular} \label{pb}}
\end{table}

\section{Stainless Steel}
\label{secss}

Stainless steel is also very often used in experiments and some
activation studies are available. Direct measurements of saturation
activity were made in LNGS \cite{lngsexposure}. Samples of stainless
steel (1.4571 grade) from different batches supplied by Nironit
company (with masses from 53 to 61~kg) were screened with GeMPI
detector at Gran Sasso for the GERDA double beta decay experiment
\cite{steel}. One of these samples was re-exposed for 314~days in
open air at the LNGS outside laboratory, after a cooling time of
327~days underground. Production rates were derived for Gran Sasso
altitude and scaled down to sea level, considering a correction
factor of 2.4. In this case, $^{60}$Co is obscured by anthropogenic
contamination, generally present in steel. The obtained results are
reported in Table~\ref{ss}.

In Ref.~\citen{mei2016}, calculations of production rates for
stainless steel have been performed using GEANT4 and ACTIVIA and are
also shown in Table~\ref{ss}. Comparing measured and calculated
rates, agreement is better in some products for ACTIVIA and in
others for GEANT4. None of the codes predicts the important
activation yield measured for $^{7}$Be.

\begin{table}
\tbl{Production rates (in kg$^{-1}$d$^{-1}$) at sea level for
isotopes induced in stainless steel, from the measurement following
an exposure to cosmic rays in Gran Sasso and from calculations using GEANT4 and ACTIVIA packages.}
{\begin{tabular}{@{}ccccc@{}} \toprule Isotope & Half-life (d)
\cite{toi},\cite{ddep} & Measurement \cite{lngsexposure} & GEANT4
\cite{mei2016} & ACTIVIA \cite{mei2016}
\\ \colrule
$^{7}$Be & 53.22 & 389$\pm$60 & 0.05 & 2.05  \\
$^{54}$Mn & 312.19 & 233$\pm$26 & 230 & 191 \\
$^{58}$Co & 70.85 & 51.8$\pm$7.8 & 90 & 13 \\
$^{56}$Co & 77.236 & 20.7$\pm$3.5 & 16 & 131\\
$^{46}$Sc & 83.787 & 19.0$\pm$3.5 & 8.8 & 18 \\
$^{48}$V & 15.9735 & 34.6$\pm$3.5 & & \\ \botrule 
\end{tabular} \label{ss}}
\end{table}

\section{Titanium}
\label{secti}

Titanium has been considered due to its properties as an alternative
to stainless steel and copper. Indeed, a reduced cosmogenic
activation is expected in this material. In the frame of the LUX
experiment, radiopurity of different titanium samples was analyzed
\cite{luxrpurity1,luxrpurity2} and the activity of cosmogenic
$^{46}$Sc was quantified, ranging from 0.2 to 23~mBq/kg. Other
scandium isotopes with shorter half-lives were also observed but are
not relevant. One 6.7~kg control titanium sample was used to
directly estimate the cosmogenic yields; the sample was screened at
the Soudan Low Background Counting Facility (SOLO) after two years
underground and then transported by ground to the Sanford Surface
Laboratory in South Dakota. The sample was activated over a
six-month period before being transported by ground back to SOLO for
re-analysis. The measured activity was (4.4$\pm$0.3)~mBq/kg of
$^{46}$Sc \cite{lux}, which agrees within a factor two with the
expected result following the calculations in Ref.~\citen{mei2016}.
In this work, production rates of $^{46}$Sc and other isotopes were
estimated using GEANT4 and ACTIVIA, as shown in Table~\ref{ti}.

\begin{table}
\tbl{Production rates (in kg$^{-1}$d$^{-1}$) at sea level for some
isotopes induced in
titanium from calculations using GEANT4 and ACTIVIA packages.}
{\begin{tabular}{@{}cccc@{}} \toprule  &
Half-life \cite{ddep} & GEANT4 \cite{mei2016} & ACTIVIA \cite{mei2016} \\
 \colrule
$^{46}$Sc & 83.787~d & 275.5 & 270.1 \\
$^{40}$K & 1.25 10$^{9}$~y & 22.1 & 61.0 \\
\botrule 
\end{tabular} \label{ti}}
\end{table}

\section{Other Results}
\label{other}

Here, some results related to the observation and analysis of
cosmogenic activation for other materials are briefly reported.

Cosmogenic activation was found to play an important role in the
scintillating {\bf CaWO$_{4}$} crystals operated as cryogenic
detectors in the CRESST-II dark matter experiment, following the
extensive background studies of a crystal (called TUM40), grown at
the Technische Universitat Munchen \cite{cresstbkg}. Distinct gamma
lines were observed in the low-energy spectrum below 80~keV,
originating from activation of W isotopes: proton capture on
$^{182}$W and $^{183}$W results in $^{179}$Ta (after a successive
decay) and $^{181}$W respectively, which decay via electron capture.
The activity of both isotopes in the crystal was quantified and, as
the leaking of events from this background source into the region of
interest for dark matter searches is very important, initiatives to
reduce the cosmogenic activation of crystals have been undertaken.
On the other hand, the production rate of tritium in CaWO$_{4}$ was
evaluated using the TALYS code \cite{mei} and it is reported in
Table~\ref{tritium}. Radioactive contamination, including some
cosmogenic products, measured for different inorganic crystal
scintillators like CaWO$_{4}$ is presented in
Ref.~\citen{damaijmpa2}.

Excitation functions of proton-induced reactions on natural {\bf
neodymium}, in the 10-30~MeV energy range, were measured in
Ref.~\citen{nd1}. Irradiation took place at the cyclotron U-120M in
Rez near Prague and the radioactivity measurement was carried out
using a HPGe detector. The measured production cross sections were
compared with TENDL-2010 predictions finding in general good
agreement in both shapes and absolute values. In addition, the
corresponding contribution to the production rates of radionuclides
relevant for double beta decay searches, like in the SNO+
experiment, was evaluated adopting the proton flux from
Ref.~\citen{barabanov}. In Ref.~\citen{nd2}, these results were
completed in 5-35~MeV energy range following an analogue
experimental strategy and comparing with excitation functions from
TENDL-2012 library.

The same experiment performed for tellurium and germanium was also
made with a natural {\bf molybdenum} target, based on 800~MeV proton
irradiation at LANSCE and screening with germanium detectors at
Berkeley \cite{norman}, for determining the $^{60}$Co production
cross section.

In Ref.~\citen{mei2016}, the production rates of a few isotopes
($^{7}$Be, $^{10}$Be and $^{14}$C) induced in {\bf PTFE} are
presented, based on analogous calculations to those performed for
xenon, copper, titanium and stainless steel using GEANT4 and
ACTIVIA.

For {\bf quartz} light guides, some production rates were calculated
in Ref.~\citen{pettus} using ACTIVIA with the Gordon et al. neutron
spectrum, including that of tritium reproduced in
Table~\ref{tritium}.

\section{Underground Activation}
\label{under}

At a depth of a few tens of meter water equivalent (m.w.e.), the
nucleonic component of the cosmic flux is reduced to a negligible
level. The neutron fluxes in underground facilities, either
radiogenic (from fission or ($\alpha$,n) reactions) or induced by
cosmic muons, in rock or set-up materials, are several orders of
magnitude lower than at surface. The spectrum of the dominant
component is concentrated at the region of a few MeV, which is below
the energy threshold of most of the spallation processes activating
materials at surface. In addition, to suppress the neutron
activation underground it could be possible to use a neutron
moderator shielding around the detector. Therefore, the cosmogenic
production underground is often assumed to be negligible. The
considered generation of radioactive isotopes underground is induced
mainly by muons. At shallow depths, the capture of negative muons is
the relevant process but deep underground interactions by fast muons
dominate (direct muon spallation or the electromagnetic and nuclear
reactions induced by secondary particles: nucleons, pions,
photons,\dots). Since the muon flux and spectrum depends on depth,
it is worth noting that underground activation studies are produced
for particular depths; therefore, comparison of different estimates
is not straightforward.

Many of the results obtained for underground activation are related
with experiments using large liquid scintillator detectors, as
summarized in the following.

\begin{itemlist}
\item An early estimate of production rates was made in Ref.~\citen{oconnel} for isotopes induced in materials typically used in neutrino experiments: C,
O and Ar. Inelastic scattering of muons giving electromagnetic
nuclear reactions was considered and rates were evaluated at sea
level and underground (2700 m.w.e.).
\item Production cross sections were measured in a reference
experiment \cite{hagner} performed at CERN irradiating with the SPS
muon beam with energies of 100 and 190~GeV different kinds of
$^{12}$C targets placed behind concrete and water to build the muon
shower like in real liquid scintillator experiments. Several
detection techniques for measuring products of different half-lives
were applied. Then, considering the measured cross-sections and the
deduced dependence with the muon energy ($\sigma\propto
E_{\mu}^{\alpha}$ with $\alpha=0.73\pm0.10$) muon induced background
rates for the muon flux at Gran Sasso and BOREXINO detector were
computed \cite{hagner}. Rates for KamLAND were estimated to be a
factor $\sim$7 higher. The most relevant contribution was that of
$^{11}$C.
\item In fact, the production rate of $^{11}$C was specifically estimated in Ref.~\citen{galbiatti} taking into account all relevant production
channels. Evaluation of cross sections from different sources
combining data and calculations was made and a FLUKA simulation of
monoenergetic muons in Borexino liquid scintillator
(trimethilbenzene) was run to derive rates and paths of secondary
particles; then, combining this information on the secondaries with
the cross sections the individual and total production rates were
derived for different muon energies. Good agreement was found with
rates coming from measurements (at 100 and 190~GeV) \cite{hagner}
and with extrapolations for average muon energies at KamLAND,
Borexino and SNOlab.
\item Analysis of KamLAND data (from 2002 to 2007) allowed the
measurement of activation yields \cite{kamland}. Isotopes were
identified and quantified using energy and time information
registered. In addition, a FLUKA simulation of monoenergetic muons
(in the 10 to 350~GeV range) was performed for KamLAND liquid
scintillator to estimate the same yields too. Comparing with
extrapolations (based on the power-law dependance with respect to
muon energy) of results from the muon beam experiment \cite{hagner},
some inconsistencies are reported for some isotopes, indicating that
estimation by extrapolation might not be sufficient.
\item In Ref.~\citen{borexino}, from data from the Borexino experiment and
profiting the large sample of cosmic muons identified and tracked by
a muon veto external to the liquid scintillator, not only the yield
of muon-induced neutrons was characterized but also the production
rates of a number of cosmogenic radioisotopes ($^{12}$N, $^{12}$B,
$^{8}$He, $^{9}$C, $^{9}$Li, $^{8}$B, $^{6}$He, $^{8}$Li, $^{11}$Be,
$^{10}$C and $^{11}$C) were measured, based on a simultaneous fit to
energy and decay time distributions. Results of the corresponding
analysis performed by the KamLAND collaboration for the Kamioka
underground laboratory are similar. All results are compared with
Monte Carlo simulation predictions using the FLUKA and Geant4
packages. General agreement between data and simulation is observed
within their uncertainties with a few exceptions; the most prominent
case is $^{11}$C yield, for which both codes return about 50\% lower
values.
\item In the context of argon-based neutrino experiments, production of $^{40}$Cl and other cosmogenically produced
nuclei (isotopes of P, S, Cl, Ar and K) which can be potential
sources of background was evaluated in Ref.~\citen{barker}.
$^{40}$Cl can be produced through stopped muon capture and the
muon-induced neutrons and protons via (n,p) and (p,n) reactions; it
is unwanted as it can be a background for different neutrino
reactions. Geant4 simulations were carried out and analytic models
were developed, using the measured muon fluxes at different levels
of the Homestake Mine. Different depths were considered in the
study, concluding that large backgrounds to the physics proposed are
expected at a depth of less than 4~km.w.e.
\end{itemlist}

Studies of underground activation have been made too in other
contexts, not related to large scintillator detectors.

\begin{itemlist}
\item Estimates of production rates for isotopes produced in enriched
germanium detectors and set-up materials (cryogenic liquid) were
made within the GERDA double beta decay experiment \cite{gerdamu},
based on a GEANT4 Monte Carlo simulation for the muon spectrum at
Gran Sasso. $^{77m}$Ge was identified as the most relevant product.

\item In Ref.~\citen{romanian}, the radioactivity induced by cosmic
rays, including neutrinos, muons and neutrons, is analyzed for the
rock salt cavern of an underground laboratory. Natural isotopes of
sodium and chlorine are considered.

\item The activation in natural tellurium due to the neutron flux
underground, in particular for Sudbury laboratory (at a depth of
about 6 km.w.e.), was analyzed in Ref.~\citen{telozza}. For
long-lived isotopes, the underground activation was estimated to be
less than 1 event/(y t). The production rates for many short-lived
isotopes of Sn, Sb and Te were quantified using ACTIVIA and TENLD
data, for neutrons coming from ($\alpha$,n) reactions as well as for
muon-induced neutrons; the obtained rates are many orders of
magnitude lower than the derived rates following the same approach
for exposure at surface.

\item The production of $^{127}$Xe in xenon detectors at the depth
of the Sanford Underground Research Facility (1480~m below surface,
4.3 km.w.e.) was evaluated in Ref.~\citen{mei2016} by Geant4
simulation, considering contributions from both fast and thermal
neutrons. A production rate of about 3~ atoms per ton per day,
considered to be negligible, was found.

\end{itemlist}

\section{Summary}
\label{sum}

Cosmogenic activation of materials can jeopardize the sensitivity of
experiments demanding ultra-low background conditions due to the
production of long-lived isotopes on the Earth's surface by nucleons
and, in some cases, also due to the continuous generation of
short-lived radionuclides deep underground by fast muons. With the
continuous increase of detector sensitivity and reduction of
primordial radioactivity of materials, this background source is
becoming more and more relevant. Direct measurements and estimates
of production rates and yields for several materials have been made
in the context of, for instance, dark matter, double beta decay and
neutrino experiments. As summarized here, cosmogenic products have
been analyzed for materials used as detector media and also for
common materials used in ancillary components. Each material
produces different relevant cosmogenic isotopes, although there are
some ones generated in most of the materials; this is the case of
tritium, which due to its decay properties is specially relevant for
dark matter searches.

In principle, sea level activation can be kept under control by
minimizing exposure at surface and applying material purification
techniques. But reliable tools to quantify the real danger of
exposing the materials to cosmic rays are desirable. There are many
options to undertake a material activation study, but unfortunately,
discrepancies between different estimates are usually
non-negligible. At the moment, there is no approach working better
than others for all targets and products. A good recipe to attempt
the calculation of the production rate of a particular isotope could
be the following:
\begin{enumerate}
\item To collect all the available information on its production cross section by neutrons and protons,
from both measurements (EXFOR database will help) and calculations,
either from libraries or using codes (based on semiempirical
formulae like YIELDX and ACTIVIA or on Monte Carlo simulation).
Measurements of cross sections at different energies using nucleon
beams are not straightforward, but they are essential to validate
models and codes giving excitation functions.

\item To choose the best description of the excitation functions of products over the whole energy range, by minimizing deviation factors between measurements and calculations.

\item To calculate the production rates considering the preferred cosmic ray spectrum and to compare them with previous estimates or measurements if available.
\end{enumerate}

Uncertainties in this kind of calculations (in many cases higher
than 50\%) come mainly from the difficulties encountered both on the
accurate description of cosmic ray spectra and on the precise
evaluation of inclusive production cross-sections in all the
relevant energy range; the low and medium energy regions below a few
hundreds of MeV are the most problematic ones since neutron data are
scarce and differences between neutron and proton cross sections may
be important. Concerning the cosmic neutron spectrum at sea level,
most of the activation studies performed in the last years have
chosen the parametrization from Gordon et al. \cite{gordon}.

Presently, the experimental results for the production rates of
cosmogenic radioisotopes are scarce and limited to a reduced number
of materials, as they require precise knowledge of the exposure
history of materials. But the availability in the future of more
measured production rates either following dedicated measurements
exposing samples to cosmic rays in a controlled environment or from
the detailed analysis of background data collected within running
experiments will help to achieve progress in the reduction of
uncertainties when estimating activation yields in materials.



\end{document}